\documentclass{article}
\input amssym.def
\input amssym.tex
\usepackage[a4paper]{geometry}
\def\ben{\begin{equation}}
\def\een{\end{equation}}

\def\bea{\begin{eqnarray}}
\def\eea{\end{eqnarray}}

\def\mathbb{\Bbb}

\newcount\hour \newcount\minute
\hour=\time  \divide \hour by 60
\minute=\time
\loop \ifnum \minute > 59 \advance \minute by -60 \repeat
\def\nowtwelve{\ifnum \hour<13 \number\hour:
                      \ifnum \minute<10 0\fi
                      \number\minute
                      \ifnum \hour<12 \ A.M.\else \ P.M.\fi
         \else \advance \hour by -12 \number\hour:
                      \ifnum \minute<10 0\fi
                      \number\minute \ P.M.\fi}
\def\nowtwentyfour{\ifnum \hour<10 0\fi
                \number\hour:
                \ifnum \minute<10 0\fi
                \number\minute}

\title{Dark Energy and Projective Symmetry}
\author{G. W. Gibbons$^1$ and  C.M. Warnick$^{1, 2}$ 
\\
\\ \small{1. D.A.M.T.P., Cambridge, Wilberforce Road, Cambridge CB3 0WA,
  U.K.}
\\ \small{2. Queens' College, Cambridge, CB3 9ET, U.K.} \\}

\begin{document}
\maketitle {\let\thefootnote\relax\footnotetext{{\em Emails}:
  g.w.gibbons@damtp.cam.ac.uk, c.m.warnick@damtp.cam.ac.uk \\ \mbox{} \hspace{.45cm}\emph{Pre-print no.} DAMTP-2010-19}}
\begin{abstract}

Nurowski [arXiv:1003.1503] has recently suggested
a link between the observation of Dark Energy in
cosmology and the projective equivalence of  
certain Friedman-Lemaitre-Robertson-Walker (FLRW) metrics.
Specifically, he points out that two FLRW metrics with the same
unparameterized geodesics have their energy densities differing by a
constant. From this he queries whether the existence of dark energy is
meaningful. We point out that physical observables in cosmology are not
projectively invariant and we relate the projective symmetry uncovered
by Nurowski to some previous work on projective equivalence in cosmology.
\end{abstract}

\section{Introduction}
In order to set Nurowski's remark \cite{Nurowski} in context we recall that
the basic physical  ideas motivating a projective viewpoint
go back to the work of Neumann \cite{Neumann}, Willam Thomson and
Peter Tait \cite{ThomsonTait}, James Thomson \cite{Thomson1,Thomson2},
Tait \cite{Tait},
Lange \cite{Lange} and  Muirhead \cite{Muirhead} 
 in the 19th century who re-examined the logical foundations
of Newtonian Mechanics, and in particular the logical status
 of Newton's first law as recounted in 
in histories of mechanics  
\cite{Mach1, Mach2,Laue,Jammer,Barbour}.
In order to give Newton's laws an operational
meaning these authors introduced the idea of an inertial frame
of reference and claimed  that the free motion
of four  particles, not moving parallel to one another,
defined a Cartesian  coordinate system such that
the motion of any other free particle would be
given by a straight line in this coordinate system.  
That is the motion of freely falling particles endows
space with a privileged set of unparameterized 
curves  providing a  projective structure and
in the absence of forces this  projective structure is flat.   

The idea  has been refined considerably since then. With
the advent of special and general relativity, attention
was transferred from space to spacetime \cite{DuVal1,DuVal2,Silberstein}.
The world lines of freely
falling particles define a projective structure on spacetime.
This corresponds mathematically to a projective connection,
i.e. an equivalence class of symmetric affine connections
all having the same {\it unparameterized}  auto-parallel curves
\ben
\Gamma _\mu \,^\nu \,_\sigma \equiv \Gamma _\mu\, ^ \nu\,_\sigma
+ A_\mu \delta ^\nu_\sigma + A_\sigma \delta ^\nu_\mu\,.
\een
However  additional physical structures, such as 
idealised clocks or light rays
are needed to justify the existence  of a metric
and relate it to the affine connection
\cite{EhlersPiraniSchild,EhlersSchild,Pfister}, and even more 
assumptions are required to obtain the dynamics of the metric. 

Weyl introduced a projective curvature tensor \footnote{not to be confused
with the Weyl conformal curvature tensor $C^\tau \,_{\mu \nu \sigma} $}  by 
\ben
W^\tau \,_{\mu \nu \sigma} := R^\tau \,_{\mu \nu \sigma} + {1 \over n-1}
\bigl(\delta ^ \tau _ \sigma R_{\mu \nu} -  
 \delta ^ \tau _ \nu  R_{ \mu \sigma  }      \bigr) \,. 
\label{Weyl} \een
and showed that a  necessary and sufficient condition for two metrics to 
share the same geodesics, considered as {\sl unparameterized } curves ,  is
that their Weyl projective tensors, with this index structure,
 coincide. It follows
that the vanishing of $ W^\tau \,_{\mu \nu \sigma}$  
is a necessary and sufficient condition for the existence
of a local coordinate chart, called a Beltrami chart,  
in which the geodesics are straight lines.
A simple calculation using  (\ref{Weyl}) then shows
that the only such ``projectively flat''
spacetimes must have constant curvature, i.e. 
they must be Minkowski, de Sitter, or Anti-de Sitter spacetime. In
Riemannian signature, this result is known as Beltrami's theorem.      
Thus these three 
spacetimes  are on an equal
footing from the purely projective point of view, just as they
are on the same footing from the purely  conformal point of view
since all three are also conformally flat. 
Thus  one can say that Newton's first law, in the form
\begin{quote} 
\it {Corpus omne perseverare in statu suo quiescendi vel movendi uniformiter in directum,
nisi quatenus a viribus impressis cogitur statum illum mutare}  
\footnote{that is:   All bodies tend to remain at rest, or to maintain a constant direction and speed,
unless forced to do otherwise, by some external force.}.
\end{quote} 
holds in any one of these three spacetimes.

At a technical level, projective techniques  
were used by Eardley and Sachs \cite{EardleySachs}
to provide spacetime with an ideal boundary:  future projective infinity.
More recently Twistorial and projective ideas 
have been applied to cosmology in   
\cite{Hurd}. However in effect Hurd makes use of the conformal
flatness of FLRW metrics rather than their truly projective properties.
In that respect his work extends to the general FLRW case,
the projective treatments  of 
de Sitter spacetime of Du Val \cite{DuVal1,DuVal2}.   

At a late stage of this investigation we were made aware of the paper of Hall and Lonie \cite{Hall} which includes some of the results below, albeit without the focus on dark energy. We would like to thank M. Dunajski for drawing this work to our attention.

\section{Dark Energy and Projective Geometry}

We now turn to Nurowski's observation \cite{Nurowski} 
and   commence  by subsuming  it  in a slightly more general result. 
This is \cite{GWW}  that a sufficient condition for two metrics
to  admit the same
unparameterized geodesics, {\sl in the same coordinates $t,x^i$}
is that if one is given by
\ben
ds^2 =-dt ^2 + a^2(t) g_{ij}(x) dx^i dx^j \,,\label{one}  
\een  
where $g_{ij}(x)$ is \emph{any} 
metric depending only on the
spatial coordinates  $x^i$, then the  other is 
\ben
ds ^2= - \frac{dt ^2}{(1-s a^2)^2 } + \frac{a^2(t)}{1-sa^2}\, g_{ij}(x) dx^i dx^j  
\label{two} \,.\een
The condition is also necessary for metrics of the form
(\ref{one}). Nurowski's result follows
by taking   
if $g_{ij}$ to  a metric of constant curvature $k$
in which both the metrics are of FLRW type.
The  more general result  follows from the calculations in 
\cite{GWW} by setting
the $f$ of that reference to $\frac{1}{a^2}$.  
The fomulation of \cite{GWW} in terms of $f$ makes it obvious 
the the projective transfomations form a one parameter symmery 
group acting as 
\ben
f \rightarrow f-s\,.
\een  
It also follows from the calculations in \cite{GWW} that
while all FLRW metrics are conformally flat, i.e.\ their Weyl tensor
vanishes, not all FLRW metrics are projectively flat.
In fact as previously noted, the only FLRW metrics
which are projectively flat are of constant curvature, that is they
must, at least locally, be isometric with de Sitter, Minkowski or
anti-de Sitter spacetime. Of course these are all (locally at least)
projectively equivalent.

If we re-write the metric (\ref{two}) as
\ben
ds^2 =-d \tau  ^2 + A^2(\tau) g_{ij}(x) dx^i dx^j \,, \label{three}  
\een  
we have 
\ben
A^2= \frac{a^2}{1-s a^2} \,,\quad  \Leftrightarrow \quad a^2  =
\frac{A^2}{1+s A^2} \,. \quad \Leftrightarrow \quad  (1-sa^2)(1+sA^2)
= 1\,. \label{Aarel}
\een
The time coordinates are related by\footnote{ 
The reader is advised that relation between the coordinate $t$
and the coordinate  $\tau$ depends on the
parameter
$s$. Thus the projective equivalence holds in the coordinates of (\ref{one}) and
(\ref{two}), but not of (\ref{three}).}  
\ben
d \tau=\frac{dt}{1-sa^2}\,,\qquad \Leftrightarrow \qquad dt=
\frac{d\tau}{1+s A^2}\,. \label{trel}
\een

Note that if $s>0$,  $a(t)$ runs from $0$ to $+\infty$  in infinite $t$ time
then $A(\tau)$ will run from  from $0$ to $+\infty$  in infinite $\tau$
time. On the other hand $A(t)$ will blow up in finite $t$ time and 
$a(\tau)$ will increase monotonically to $\frac{1}{s}$ in infinite
$\tau$ time. 

Differentiating the relation of (\ref{Aarel}) and making use of
(\ref{trel}) we deduce that
\ben
\frac{1}{a} \frac{da}{dt} = \frac{1}{A} \frac{dA}{d\tau}.\label{inv1}
\een
Noting that (\ref{trel}) implies
\ben
a^2 \frac{d}{dt} = A^2 \frac{d}{d\tau} \,,
\een
we can further differentiate to find
\begin{eqnarray}
a \frac{d^2 a}{dt^2} - \left(\frac{da}{dt} \right)^2 &=& A \frac{d^2
  A}{d\tau^2} - \left(\frac{dA}{d\tau} \right)^2  \label{inv2}\\
a^3 \frac{d^3 a}{dt^3}- a^2\frac{d^2 a}{dt^2}\frac{da}{dt} &=& A^3 \frac{d^3 A}{d\tau^3}- A^2\frac{d^2 A}{d\tau^2}\frac{dA}{d\tau}.\label{inv3}
\end{eqnarray}
Clearly these expressions are invariant under the projective
transformation givexconstruct invariants of higher
order in time derivatives. The second order
invariant (\ref{inv2}) appears in the components of the projective
curvature of the metric (\ref{two}). We find therefore that the Hubble
expansion rate  $H= \frac{1}{a} \frac{d a }{dt } $  is projectively invariant, 
however the deceleration 
$ q=- a \bigl(\frac{d a }{dt } \bigr)^{-2} \frac{d^2 a }{dt ^2}   $  and 
jerk $j= a^2 \bigl(\frac{d a }{dt } \bigr)^{-3} \frac{d^3 a }{dt ^3} $ 
 \cite{GibbonsDunajski}
are not. The statements
\ben
q=1, \quad \Leftrightarrow \quad a(t) = c_1 e^{c_2 t},
\een
and
\ben
j=-q \quad \Leftrightarrow \quad a(t) = c_1 \cos c_2 (t-t_0), \quad \textrm{or}
\quad a(t) = c_1 \cosh c_2 (t-t_0),
\een
are projectively invariant and are obeyed by the projectively flat
Minkowski and de Sitter metrics. In general these metrics may be
characterised as having an energy density $\rho = c + d/a^2$
where $c$ and $d$ are constants.

If we assume the spatial metric to be of constant curvature then the
Einstein equations imply the Friedman equations
\bea
\frac{{\dot a}^2 }{a^2 } + \frac{k}{a^2} &=& \frac{8 \pi}{3} \rho \\
\frac{{A^\prime}^2}{A^2} + \frac{k}{A^2} &=& \frac{8 \pi }{3} \tilde
\rho 
\eea
where $\rho$ is the energy density of metric (\ref{one}) and $\tilde
\rho$
that of metric (\ref{two}). 
A simple calculation, making use of (\ref{inv1}), reveals the remarkable fact that
\ben
\tilde \rho= \rho - \frac{3ks}{8 \pi} \,. \label{cosmic}
\een
A brief calculation making use of the conservation equation shows that
the pressure transforms according to
\ben
\tilde p = \frac{p-sA^2 \rho}{1+sA^2} + \frac{3 k s}{8 \pi}.
\een

The relation (\ref{cosmic}) suggests that a metric sourced by a pure  Cosmological
Constant $\Lambda$  is projectively equivalent under the
transformation above to one with a Cosmological Constant $\tilde \Lambda$ given by 
\ben
\tilde \Lambda = \Lambda -  3ks \,. \label{lam}
\een
This can be seen explicitly. For example, we may take the $k=1$ metric
of de Sitter
\ben
ds^2 = -dt^2 + l^2 \cosh ^2 (t/l) d\Omega_3^2
\een
where $d\Omega_3^2$ is the metric on the unit $3$-sphere. This is the
metric for the de Sitter space of constant curvature $\Lambda=3/l^2$. From
(\ref{trel}) we deduce that
\ben
\frac{1}{l} \tanh \frac{t}{l} =  \frac{1}{l'} \tanh \frac{\tau}{l'},
\een
where $(l')^{-2} = l^{-2} -s$. Making use of this, we find the
identity
\ben
A^2 = \frac{l^2 \cosh^2 (t/l)}{1-s l^2 \cosh^2 (t/l)} = (l')^2 \cosh^2 (\tau/l'),
\een
so that the new metric is again de Sitter with a $k=1$ slicing, but
with a new cosmological constant $\Lambda' = 3/(l')^2 = \Lambda -
3ks$. Similar calculations verify (\ref{lam}) directly in the case of de Sitter
with $k=-1$ and anti-de Sitter with $k=-1$. For the $k=0$ slicings of
de Sitter one
finds that the transformation given above leads to exactly the same form of
the metric. In this instance, the projective transformation is simply a
diffeomorphism.

One might think that (\ref{cosmic}) implies that the projective transformation
is equivalent to changing the matter content of the universe by adding
a cosmological constant. It should be borne in mind, however, that the equation of state
of the matter is not in general invariant under  the  projective transformations,
because in in the  FLRW metric (\ref{one}) $\rho$ depends on $a$
while in the projectively equivalent metric (\ref{two}) 
it depends on  $A$.  Thus for example if metric (\ref{one}) is
supported
by matter with negligible pressure (\lq \lq dust \rq \rq) 
then 
\ben
\rho= \frac{C}{a^3} \,, \qquad \Rightarrow \qquad  \Lambda =0 
\een  
and so
\begin{eqnarray}
\tilde{\rho} &=& 
  \frac{C}{A^3} \Bigl( 1+s A^2 \Bigr )^{\frac{3}{2}}
 -\frac{3ks}{8\pi}, \\
\tilde{p} &=&  \frac{3ks}{8\pi} - C\frac{s}{A}\left( 1+s A^2\right)^{\frac{1}{2}}
\end{eqnarray}
This differs from the result of Nurowski, who separates out a
time-dependent cosmological constant in order that the matter content
of the transformed metric take the form of dust with a cosmological
constant. Since such a separation results in an energy-momentum tensor
which is not conserved, we prefer to leave the energy-momentum intact.

For small $A$, i.e. for small $\tau$ time, 
\ben
\tilde \Lambda \approx 3ks\,,
\een
but at late $\tau$  time
\ben
\tilde \Lambda \approx -3ks +8 \pi C s^\frac{3}{2}.
\een
in other words there is a  effective 
$\tau$-time dependent
Cosmological \lq \lq Constant \rq \rq as judged by the projectively 
equivalent metric (\ref{two}).

\section{Physical Observables}

If
\ben
ds ^2 = -dt^2 + a^2(t) \Bigl \{ dr ^2 + f^2_k(r) \bigl( d \theta ^2 + \sin ^2 \theta d \phi ^2 \bigr ) \Bigr \}\,,
\een
with
\ben
f_k(r)= \sin r\,,r\,,\sinh r \,,\qquad {\rm if } \qquad k= 1\,,0\,,-1\,,
\label{sine}\een
then a  source emitting  at $t=t_e$ is observed  at $t=t_o$
with redshift
\ben
1+z= \frac{a(t_o)}{a(t_e)} \label{redshift} \,.
\een
Clearly the redshift is a physically measurable
observable which is not invariant under the
projective transformations (\ref{Aarel}).    

The difference  in the radial coordinate of source   and
observer is    
\ben
r(t_o) - r(t_e)  = r=  \int ^{a(t_o)} _{a(t_e)}
 \frac{ da}{\sqrt{\frac{8 \pi }{3} \rho (a) a^4 -k a^4  } }\,. \label{radius} 
\een     
The intensity $I(t_0)$ received is
\ben
I(t_o) = \frac{L(t_e) }{4 \pi} \frac{1}{ f_k^2(r) (1+z)^2} \,, 
\label{luminosity} \een
where $L(t_e)$ is the total emitted luminosity.
In astronomy it is traditional to define the magnitude $m(t_o)$  a
source  by the logarithm of the received intensity $I(t_o)$. 
Using equations   (\ref{sine})-(\ref{luminosity})
it one may construct a curve of $m$ against $z$ which is
the redshift magnitude relation. This relation forms a major part of
the evidence for the existence of dark energy. It is not  invariant
under projective transformations.

The angular size $\delta \phi$ of an sources of proper size $l_e$ is
\ben
\delta \phi= \frac{l_e}{a(t_e)f_k(r)} \,,
\een
and so similar remarks apply to the angular diameter redshift relation.

\section{Conclusion}

As emphasised by Ehlers Pirani and Schild \cite{EhlersPiraniSchild},
spacetime  requires more than
just a projective structure if it is to describe the real
world, since  we can do more than just observe
the (unparameterized) paths of freely falling particles.
We can observe the motion of light rays, which determines
a conformal structure and  we can also measure, using atomic clocks,
the lapse 
of proper time along the world lines of particles.
This provides us with a preferred parameterization  
and in  effect  determines a pseudo-Riemannian structure.
In the case of FLRW cosmology 
the metric structure is determined from, amongst other things,
the observed  redshift magnitude relation.
As we have seen, this is not projectively invariant.
The existence of dark energy is inferred, among other things, from
the redshift magnitude relation, and this is also not
projectively invariant.  

Interestingly, the situation resembles a recent debate
concerning  whether the cosmological constant  $\Lambda$ 
affects light bending,
since the (unparameterized) projections of light rays in 
the static  Schwarzschild-de Sitter
or Kottler metric onto the spatial manifold do not depend on $\Lambda$
(see \cite{GWW} and references therein).
The origin of this phenomenon is that the projective structure of the
associated optical metric is independent of $\Lambda$ \cite{GWW}. However
this does not imply that there is no effect on observations
of gravitational lensing since that involves measuring angles
which are not invariant under  projective transformations
of the optical metric \cite{GWW}. 

Nevertheless, while disagreeing with the suggestion   \cite{Nurowski} that
``dark energy is meaningless'' we believe that 
some  intriguing connections have been uncovered 
which may conceivably lead to  a deeper
understanding of dark energy.
It is striking that the reciprocal relations
or dualities 
(\ref{Aarel}) recall those arising other contexts, such as 
 Born-Infeld theory. Another interesting point relates
to  an old observation (see \cite{GibbonsDunajski} and references therein)
that much  present day cosmological observations may be 
summarised by the statement that the universe is
of FLRW type with jerk $j=1$. This implies that $k=0$
and that the universe was pressure free matter dominated at early times and
dark energy dominated at late times. As we have seen, this is not
a projectively invariant statement. But as we have also seen, the
relations $q=-1$ and $j=-q$ are projectively invariant and are
satisfied by solutions containing a cosmological constant and curvature energy.

\end{document}